\documentclass[a4paper,12pt]{article}

\usepackage [dvips]{graphicx}
\usepackage{amsmath,amssymb,color,ulem} 

\begin{document}

\title{Length regulation of microtubules  by molecular motors: \\ Exact solution and density profiles}

\author{Chikashi Arita, Alexander L\"uck\footnote{
Present address: Informatik, Universit\"at des Saarlandes, 66041 Saarbr\"ucken, Germany} , Ludger Santen \\
\normalsize Theoretische Physik, Universit\"at des Saarlandes, 66041 Saarbr\"ucken, Germany} 

\date{}

\maketitle

\begin{abstract}
In this work  we study a microtubule (MT) model, whose length is regulated by the action of processive kinesin motors. We treat the case of infinite processivity, i.e. particle exchange in the bulk is neglected.  The exact results can be obtained for model parameters which correspond to a finite length of the MT. In contrast to the model with particle exchange we find that the lengths of the MT are exponentially distributed in this parameter regime. The remaining parameter space of the model, which corresponds to diverging MT lengths, is analyzed by means of extensive Monte Carlo simulations and a macroscopic approach. For divergent MTs we find a complex structure of the phase diagram
 in terms of shapes of the density  profile. 
\end{abstract}

\section{Introduction}
The cytoskeleton is a dynamic network of biopolymers which determines the shape of eukaryotic cells and serves at the same time as a transport network. One distinguishes three different types of cytoskeletal filaments (actin, microtubules and intermediate filaments). Microtubules (MTs) consist of  $\alpha$ and $\beta$ tubulin subunits and form cylindrical structures with high bending rigidity \cite{bib:Alb}. The dynamic properties of MTs allow  them to take on arbitrary length-distributions, which are adapted to the different cell types and shapes. Therefore it is of fundamental interest to identify the mechanisms, which control the length distributions of MTs. These regulatory mechanisms are carried out by polymerases, which support  MT growth,  or by depolymerases, which are able to induce fast depolymerization events, so-called catastrophes \cite{bib:GZH}.  The stabilization of growing MTs has strong influence on their length distribution, in particular under spatial confinement \cite{bib:EbbiS}. Dynamically stabilized MTs are able to connect the cell center and membrane without losing their ability to adapt to different cell shapes. Other proteins have been identified which are able to enhance the depolymerization of MTs. This is the case for Kip3 a member of the kinesin-8 family with extremely high processivity. Kip3 moves to the plus end of MTs where it is able to enhance the depolymerization activity. This process is of great importance in order to position the mitotic spindle in yeast  \cite{bib:GCRP,bib:VHTHTH,bib:VLBDH}.

The length regulation mechanism induced by Kip3 has been addressed theoretically  in  \cite{bib:RMF,bib:JEK,bib:MRF,bib:RMFa}, by using one-dimensional driven diffusive processes. These models capture the stepwise directed motion of the kinesins and describe their interaction by mutual exclusion. 

From a theoretical point of view driven diffusive processes are stochastic models of interacting  particles, which have been studied  typically on  lattices with a fixed system length  (i.e. with a fixed number of sites)   or on $\mathbb Z $ \cite{bib:Schu}.  The totally asymmetric simple exclusion processes (TASEP) is the archetype of  this kind of models.  One of well-studied cases is the TASEP with open boundaries,    i.e.  on a finite chain, particles are injected at the one end and extracted at the other end of the chain \cite{bib:MG,bib:DEHP}. We refer to this realization of the model as open TASEP.  Recently the TASEP or its related processes with varying system size have been intensively studied, motivated by other biological applications or modeling queues with exclusion principle \cite{bib:SEPR,bib:SE,bib:EvanS,bib:DMP,bib:Ari,bib:AS,bib:dGF} as well as the MT models  \cite{bib:RMF,bib:JEK,bib:MRF,bib:RMFa,bib:RMFb}.

In this work we study a variant of the TASEP, which was introduced in \cite{bib:MRF} as a ``simplified''  model of length regulation via kinesin motors. The parameters characterizing the model are the motor protein input rate, the polymerization rate and  the depolymerization rate. Exchange of motors in the bulk, i.e. Langmuir kinetics, is not considered, which seems to be justified in view of the high processivity of Kip3.  (Langmuir kinetics was also introduced in the full model \cite{bib:MRF,bib:JEK}.) 

\begin{figure}[t]
\includegraphics[width=16cm]{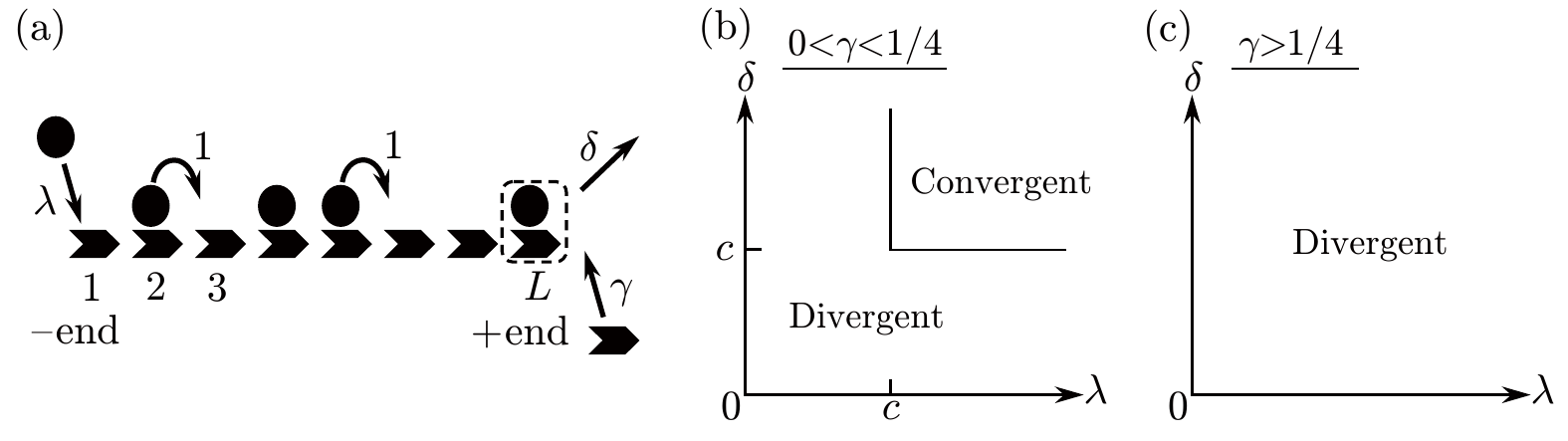}
\caption{(a) Schematic illustration of the model dynamics: Each binding site of the MT, which is represented as a lattice site, is occupied by a single motor or empty.  New motors are injected only at the $-$ end, which can be regarded as reservoir with density $\lambda$.  At the $+$ end, polymerization and depolymerization occurs. We assume that the depolymerization is allowed only if the last site is occupied by a motor.  (b,c) The phase diagram in terms of converging MTs and diverging MTs  ($c = \frac{1-\sqrt{1-4\gamma}} 2$):   The convergent phase appears when $\gamma<\frac 1 4$ (b).  For $\gamma > \frac 1 4$ (c), MTs always diverge. 
 \label{fig:schema_phase}}
\end{figure}

We denote  the  length of the MT by $L$, where each site is labeled by an integer $j$ from the minus end $j=1$ to the plus end $j=L$, see Fig.~\ref{fig:schema_phase} (a). Molecular motors occupy a single site and interact via mutual exclusion. We write $\tau_j=1,0$, if the site $j$ is occupied or unoccupied, respectively. Therefore the state space $S$  consists of configurations $\tau=\tau_1\cdots \tau_L(L\ge0)$, where we write $\tau=\emptyset$ for $L=0$. A motor at site $j$   walks  to site $j+1$ with rate 1, if site $j+1$ is not occupied by another motor.  A new motor enters the system with rate $\lambda$ (in \cite{bib:MRF},  this is denoted by $\rho_-$) at the minus end if the first site exists and is not occupied. 

In contrast to the open TASEP the lattice size which represents the length of the MT,  is not fixed.  At the plus end of the MT, polymerization occurs, i.e. a new site is attached to the lattice with rate $\gamma$.  If the MT is completely depolymerized, i.e. $L=0$, a new site is {\it created} with rate $\gamma$.  Depolymerization occurs with rate $\delta$,  if the site $L$ is occupied. In this case the last site is removed together with the particle that occupies this  site.   Note that we do not take into account the cell size, i.e. no upper bound of the MT length $L$ is introduced. 

In this work we present an exact analytical approach to the stationary properties of the model for finite MT lengths. We also use a phenomenological approach and Monte Carlo simulations for the case of diverging MTs.  We show in section \ref{sec:stationary} (and in  \ref{sec:proof} for details) that the stationary measure can be written in a matrix product form, which is a ``grand canonical'' version of the open TASEP.  By evaluating the normalizability of the stationary measure, we determine the area in the parameter space where the MT length converges. In this case, the average MT length can be exactly calculated by using some known results from the open TASEP.  In order to analyze the divergent case in section \ref{sec:divergent-phase} we  give some analytical arguments: For divergent MTs the macroscopic density profiles can be of shock or rarefaction type, depending on the  parameters of the model.   We also check by Monte Carlo simulations that the predicted density profiles are actually realized.  In Section \ref{sec:discuss}, we summarize this work and give some concluding remarks. 

\section{Stationary distribution}\label{sec:stationary}
Depending on the growth rate $\gamma$, one observes a stationary length distribution of the MTs ($\gamma < \gamma_c$) or a divergent MT length ($\gamma > \gamma_c$).  In \cite{bib:MRF}  the critical rate $\gamma_c$ has been established for different input and depolymerization rates   
\begin{align}\label{eq:gamma_c}
   \gamma_c = \left\{ \begin{array}{ll}
   \lambda (1-\lambda)   & (\lambda<\frac 1 2 ,\lambda<\delta ),  \\
   \delta (1-\delta )   & (\delta<\frac 1 2 ,\lambda\ge \delta ), \\
   \frac 1 4            & (\lambda,\delta\ge \frac 1 2 ) . 
    \end{array}\right.
\end{align}
Figure \ref{fig:schema_phase} (b,c) shows the phase diagram in $\lambda$-$\delta$ plane  for given $\gamma$, which is equivalent to this result.  The convergent phase is the region where both $ \delta  $ and $  \lambda $   are larger than the critical value  $ c = \frac{1-\sqrt{1-4\gamma}}{2}  $   (a solution to  $c(1-c)=\gamma$)  for $ 0<\gamma <\frac 1 4  $.  

The form of $\gamma_c$ is mathematically the same as the  stationary current of the open TASEP with fixed lattice size, where particles are injected with rate $\lambda$ and extracted with rate $\delta$  \cite{bib:MG,bib:DEHP}. In \cite{bib:DEHP} it has been shown that the stationary weights of the open TASEP on static lattices can be represented by (in general) infinite dimensional  matrices $X_1,X_0$ and vectors $\langle W| ,|V\rangle$ that satisfy the following algebra: 
\begin{align}
 \label{eq:algebra-bulk}  X_1X_0 & =X_1+X_0,    \\
 \label{eq:algebra-left} \lambda \langle W| X_0 & = \langle W|,     \\
 \label{eq:algebra-right} \delta X_1 |V\rangle  & = |V\rangle. 
\end{align}
For a given system length $L$, 
the stationary probability of finding  a configuration  $ \tau_1 \cdots \tau_L $ is represented as 
\cite{bib:DEHP} 
\begin{align}
   P (\tau_1 \cdots \tau_L) = \frac {1 }{ Z_L }
   \langle W|  X_{\tau_1} \cdots X_{\tau_L}    |V\rangle 
\end{align}
with the  normalization 
\begin{align}\label{eq:ZL=}
   Z_L  =  \sum_{\tau_j =   1 ,0}  \langle W|  X_{\tau_1} \cdots X_{\tau_L}    |V\rangle
    = \langle W| ( X_1+ X_0 )^L    |V\rangle. 
\end{align}
 From (\ref{eq:algebra-bulk}) it is evident that the stationary current is calculated as 
\begin{align}
 J= \langle \tau_i (1- \tau_{i+1} ) \rangle
 =  \frac{1}{Z_L} \langle W| ( X_1+ X_0 )^{i-1} X_1 X_0   ( X_1+ X_0 )^{L-i-1}  |V\rangle 
 = \frac{ Z_{L-1} }{ Z_L }   .
\end{align} 
 In \cite{bib:DEHP} it was also shown  that the stationary current approaches the values described  in (\ref{eq:gamma_c})  as $ L\to \infty $. 

Let us go back to the MT model.  We note that, in contrast to the open TASEP, the MT length $L$ is also a variable.  A ``grand canonical ensemble'' of the open TASEP gives a stationary measure of this process, where the polymerization rate $\gamma$ plays the role of fugacity \cite{bib:L}\footnote{Here $ \gamma^L $ corresponds to the states  where the number of lattice sites is $L$ (not the number of particles).  When the polymerization rate  $\gamma_L$ depends on the MT length $L$, a simple generalization of (\ref{eq:stationary-measure}) with replacement  $\gamma^L \to \gamma_0\gamma_1 \cdots \gamma_{L-1} $  describes the stationary measure.}  (see \ref{sec:proof} for details): 
\begin{align}\label{eq:stationary-measure}
   F (\emptyset) = 1, \ 
   F (\tau_1\cdots \tau_L) = \gamma^L 
   \langle W|  X_{\tau_1} \cdots X_{\tau_L}    |V\rangle.
\end{align}
A similar realization of the grand canonical ensemble of the open TASEP  has been previously found in \cite{bib:H}. Polymerization and depolymerization  at the right boundary lead  to mathematically the same condition as in \cite{bib:H}\footnote{In the model of \cite{bib:H}, the left end is also dynamical, where a site occupied by a particle is attached.  When the leftmost site is empty, it is removed from the system.}.    This recipe has been also applied in order to construct an exact stationary solution to a queueing process with exclusion \cite{bib:Ari},  where the arrival rate plays the role of fugacity. However, constructing a matrix product form with some parameter as fugacity is not always possible. One counterexample is given by another queuing process  with high and low priorities of customers  \cite{bib:dGF}, where it seems difficult to find a matrix product solution although this model can be regarded as an exclusion process with varying length.

The form (\ref{eq:gamma_c}) is a direct consequence of the normalizability of the stationary measure.  In order to check for convergence of the normalization $ \mathcal Z  $,  we first note that  the stationary measure for a given length $L$ is  given by
 \begin{align}
     \sum_{\tau\in S \atop (\text{length}=L)} F(\tau)=  
     \sum_{\tau_j=1,0  }\gamma^L    \langle W|  X_{\tau_1} \cdots X_{\tau_L}    |V\rangle
        = \gamma^L Z_L . 
 \end{align}
The normalization of the open TASEP $Z_L $ was calculated as \cite{bib:DEHP}
\begin{align}  
Z_L   = \sum_{j=1}^{L+1}\frac{j(2L-j)!}{L!(L-j+1)!}
   \frac{ \lambda^{-j}-\delta^{-j}}{1/\lambda-1/\delta}     
  \sim 
   \left\{\begin{array}{ll}
     \big[ \lambda (1-\lambda) \big]^{-L}   & (\lambda<\frac 1 2 ,\lambda<\delta ),  \\
     \big[ \delta (1-\delta ) \big]^{-L}  & (\delta<\frac 1 2 ,\lambda\ge \delta ), \\
    4^L             & (\lambda,\delta\ge \frac 1 2 ) .
    \end{array}\right.
    \label{eq:z_asym}
\end{align}
From this behavior  we find that 
\begin{align}
\mathcal Z =  \sum_{\tau\in S }   F (\tau)
= \sum_{L\ge 0}  \gamma^L Z_L 
\end{align} 
is finite when $\gamma<\gamma_c$ with equation~(\ref{eq:gamma_c}). In other words, for $\gamma<\gamma_c$ 
 the stationary measure (\ref{eq:stationary-measure}) is normalizable  and therefore  $\frac{1}{\mathcal Z } F (\tau)$ gives the exact stationary distribution. The critical value $\gamma_c$ corresponds the case where $ \lim_{L\to \infty } Z_L \gamma_c^L= 1  $.

\begin{figure}
\includegraphics[width=16cm]{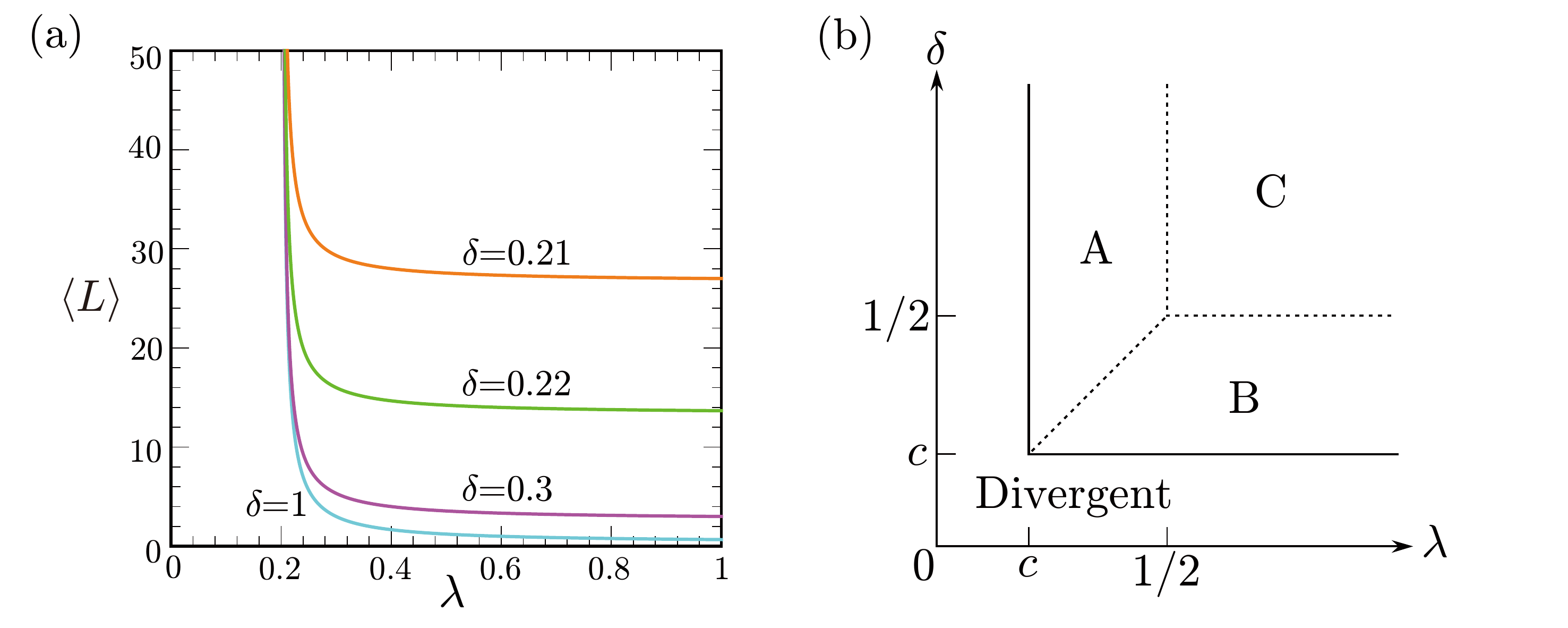}
\caption{(a) Average MT length $ \langle L\rangle  $ (\ref{eq:<L>}) vs $\lambda$ in the stationary distribution. The values of $\delta$ are indicated in the panel, and $\gamma=0.16$. The average length diverges as $\lambda$ approaches  the critical value $ c= \frac{1-\sqrt{1-4\gamma}}{2}=0.2$.  (b) Subphases of the convergent phase. In the three regions A, B and C,  the length distribution exhibits the different asymptotic  behaviors as Equation  (\ref{eq:P=}). }
\label{fig:<L>}
\end{figure}

The grand canonical partition function of the open TASEP is known to have the form \cite{bib:BJHK,bib:BE} 
\begin{align}\label{bib:mathcalZ=}
  \mathcal Z = \frac{4\lambda\delta}{
    ( \sqrt{1-4\gamma} -1 + 2\lambda)( \sqrt{1-4\gamma} -1 +2\delta)}  
\end{align}
for $\gamma <\gamma_c $. Furthermore, the stationary distribution of the length $L$ is given by $\gamma^L Z_L/\mathcal Z$, and its mean value converges to 
\begin{align}
 \langle L\rangle  = \sum_{L\ge 1 } L \frac{  \gamma^L Z_L}{\mathcal Z}  
  = \gamma \frac{\partial}{\partial\gamma}\log\mathcal Z 
  = \frac{4\gamma(\sqrt{1-4\gamma}-1+\lambda+\delta)}{\sqrt{1-4\gamma}
    ( \sqrt{1-4\gamma}-1 +2\lambda)(\sqrt{1-4\gamma}-1+2\delta)}  .
    \label{eq:<L>}
\end{align}
Except for the cases where $\lambda$ is close to the phase boundary $\lambda = c $, the average length does not take a large value (typically e.g. $\langle L\rangle<10$), as shown in Fig.~\ref{fig:<L>}  (a).  Moreover, the lengths $L$ are (except for very small values)  exponentially distributed,  as one can gather from  (\ref{eq:z_asym}): 
\begin{align} 
 \label{eq:P=}
 P(L)
 = \gamma^LZ_L/\mathcal Z
  \sim \left\{\begin{array}{ll}  
   \Big[\frac{\gamma}{\lambda (1-\lambda)} \Big]^L   &
    ( \text{A: }\  c <  \lambda<\frac 1 2 ,\lambda<\delta ),  \\
   \Big[\frac{\gamma}{\delta (1-\delta )} \Big]^L    &
    ( \text{B: }\  c < \delta<\frac 1 2 ,\lambda\ge \delta ), \\
    (4\gamma)^L             & 
    ( \text{C: }\  \lambda\ge \frac 1 2 ,\delta\ge \frac 1 2 ) .
     \end{array} \right.
\end{align}
 One can distinguish three subphases in the convergent phase as shown in Fig. \ref{fig:<L>} (b), according to the asymptotic behavior of the length distribution.  

The formula (\ref{eq:<L>})  differs from Equation~(8) of \cite{bib:MRF}, where the Langmuir dynamics in the bulk is taken into account. These results underscore the importance of Langmuir dynamics for length regulation of MTs in infinite volume. 

The inflow and outflow of lattice sites should be balanced in the stationary distribution. In other words,  the ``tip density''  $\rho_+  = \langle\tau_L\rangle$ must satisfy  $ \gamma = \delta \rho_+   $. The matrix product solution is compatible with this relation. In fact we have
\begin{align}\label{eq:stat-dist-rho_+}
\rho_+ = \frac{1}{ \mathcal Z } 
\sum_{L\ge 1 } \gamma^L\langle W | (X_1+X_0)^{L-1} X_1 |V\rangle 
=  \frac  \gamma  \delta   
\end{align} 
thanks to  relation (\ref{eq:algebra-right}).

In this section, the matrix product form enabled us to show the  condition $\gamma <\gamma_c  $ with  (\ref{eq:gamma_c}) for the convergence of the MT length, and to calculate the length distribution in the convergent phase.   However, it does not give a rigorous result  for $\gamma>\gamma_c$,    since in this case no stationary state is reached and   the length of the MT diverges.  This implies for example that (\ref{eq:stat-dist-rho_+}) is no longer  the  correct tip density when the MT length diverges.  Thus we shall rely on Monte Carlo simulations and a phenomenological approach to find density profiles in the divergent phase in the next section. 

\section{Divergent phase}\label{sec:divergent-phase}

In this section we will characterize the states in the divergent phase on a macroscopic level, by using the conservation of mass.  
\begin{figure}
\includegraphics[width=16cm]{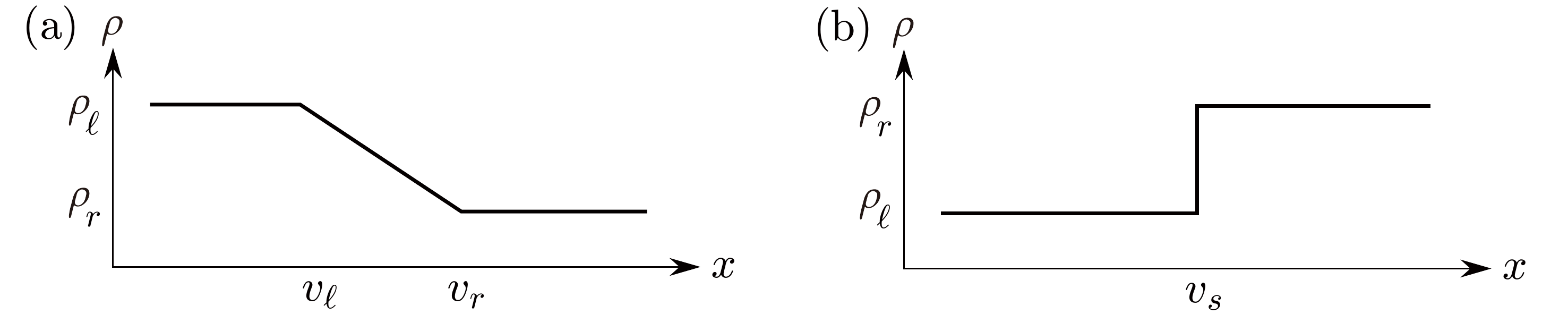}
\caption{Macroscopic density profiles for particles that move in positive $x$-direction. (a) Particles move from a high density domain ($\rho_r$)  toward  a low density domain ($\rho_\ell$). A  rarefaction wave evolves. The left (right) end of the slope part is moving with  velocity $v_\ell=1-2\rho_\ell$ ($v_r=1-2\rho_r$).  (b) A shock wave is established if particles move from a low toward a high density domain.}
\label{fig:rs}
\end{figure}

Macroscopically we can identify domains of different density $\rho(x)$, in which the current $J(x)$ is given by $J(x)=\rho(x)(1-\rho(x))$.  In general the current in different domains may differ, which means that mass is transferred from one domain to another. Let us consider the typical two waves appearing in the TASEP \cite{bib:KRB}. If mass is transferred from a high to a low density domain, this leads to rarefaction density profiles 
\begin{align}\label{eq:rare}
    \rho(x)= 
    \left\{\begin{array}{ll}
       \rho_\ell & (x<1-2\rho_\ell) \\
       \frac 1 2 ( 1-x ) & (1-2\rho_\ell<x<1-2\rho_r) \\
       \rho_r & (x>1-2\rho_r) 
     \end{array} \right.
\end{align}
with $\rho_\ell>\rho_r$, see Fig.~\ref{fig:rs} (a). Shock wave density profiles
\begin{align}\label{eq:shock}
    \rho(x) =  
    \left\{\begin{array}{ll}
      \rho_\ell  & (x<v_s=1-\rho_\ell-\rho_r) \\
       \rho_r & (x >v_s) 
     \end{array} \right.
\end{align}
are observed if  $\rho_\ell<\rho_r$, see Fig.~\ref{fig:rs} (b). The rarefaction and shock wave positions are stationary in the frame in which the spatial position $j$  is scaled by the time $t$ as $x=j/t$. 

In earlier works \cite{bib:AS} and \cite{bib:dGF}, density profiles of the queueing processes with exclusion were analyzed. It turned out that the macroscopic density profiles can be  rarefaction or shock type, respectively.  We shall show that both types are  also realized in the MT model, depending on the parameter values.  

We are interested in the macroscopic density profile of the MT $\rho(x)\ (0<x<v_+)$, where $v_+$ denotes  the velocity of the MT's plus end.  For our analysis it is important to notice  that the actual  density $\rho_+$ at the plus end is, in general, different from the density $R$ of the macroscopic density profile at the  plus end of the MT:
\begin{align}
\label{eq:R:=}
R := \lim_{x\nearrow v_+ }\rho(x)   .
\end{align}
 The tip velocity and density are related to each other via 
\begin{align} \label{eq:v_+=}
    v_+ = \gamma-\delta \rho_+   
\end{align}
(\textit{cf.} Equation~(1) of \cite{bib:MRF}), since the effective depolymerization rate of the MT is given by $\delta \rho_+$, which is identical to particle current in the frame of the tip, i.e. in the co-moving frame. Therefore, the particle conservation law in the co-moving frame is expressed  as\footnote{This is equivalent to Equation~(2) in \cite{bib:MRF}, where $R$ is denoted by $\rho_b$.   }
\begin{align}\label{eq:conservation}
    R(1-R)- R ( \gamma -\delta \rho_+ )= \delta \rho_+,
\end{align}
where the first term denotes the current generated by particle hopping,  the second term is the current induced by moving the frame. 
In \cite{bib:MRF}, subphases in the divergent phase, i.e. EX,  MC  and IN phases, were found according to the form of $ \rho_+ $. Here we investigate which kind of macroscopic density profiles can be observed in the subphases,
 and we inversely determine the phase boundaries by considering
 conditions for existence of the density profiles.

\subsection{EX phase}

Let us begin with the assumption that the tip density $ \rho_+ $ \textit{prefers}  to become identical to the right end of the macroscopic density $R$. In this case we have 
\begin{align}\label{eq:rho+EX}
   \rho_+ = 1- \frac{\gamma}{1-\delta}  
\end{align}
due to the particle conservation (\ref{eq:conservation}) with  $\rho_+=R $.  This case corresponds to the EX phase as labelled in \cite{bib:MRF}. The macroscopic density profile is not always simply flat, which was not emphasized previously.  It can become a rarefaction wave (\ref{eq:rare}) or a shock wave (\ref{eq:shock}) with left and right domain densities as $\rho_\ell = \lambda $ and $ \rho_r =  \rho_+ $. In the EX phase there are in general four types of macroscopic density profiles:
\begin{align} 
\label{eq:p-s-p}
\rho(x)\big|_{ \text{EX-I}} & = 
\left\{\begin{array}{ll}
     \lambda & ( 0<x <1-2\lambda), \\
      \frac{1-x}{2} &
       (  1-2\lambda<x<1-2\rho_+), \\
     \rho_+  &  (1-2\rho_+<x<v_+) ,
  \end{array} \right.   
\\   \rho(x)\big|_{ \text{EX-II}} & = 
\left\{\begin{array}{ll}
      \frac{1-x}{2} &
       (0<x<1-2\rho_+) ,\\   \rho_+&  (1-2\rho_+<x<v_+) ,
  \end{array} \right.   
\\
\rho(x)\big|_{ \text{EX-III}} & =  \rho_+   \   (0<x<v_+) ,
     \label{eq:single}
\\
\label{eq:p-p}
\rho(x)\big|_{ \text{EX-IV}} & = 
 \left\{\begin{array}{ll}
     \lambda & ( 0<x <v_s) \\
    \rho_+ & ( v_s<x<v_+) ,
  \end{array} \right.   
\end{align}
with the shock and tip velocities 
\begin{align}
v_s=   \frac{\gamma}{1-\delta} -\lambda  , \quad 
v_+= \gamma - \delta (1- \frac{\gamma}{1-\delta}).
\label{eq:v-v-EX}
\end{align}
Because of the definition of the EX phase (\ref{eq:rho+EX}), the boundaries between this subphase and the other two subphases  can also be given analytically by simple considerations:  The transitions  from the EX phase to the others occur when the relative size of the density domain $ \rho(x) = \rho_+  $  shrinks   i.e.   $ 1-2\rho_+ \leq v_+  $  (in the case of the rarefaction wave)    or  $v_s \leq v_+$  (in the case of the shock wave).  We obtain  $ \delta = 1-\sqrt\gamma $ and $\lambda = \delta$,  respectively.
  
\begin{figure}
\includegraphics[width=16cm]{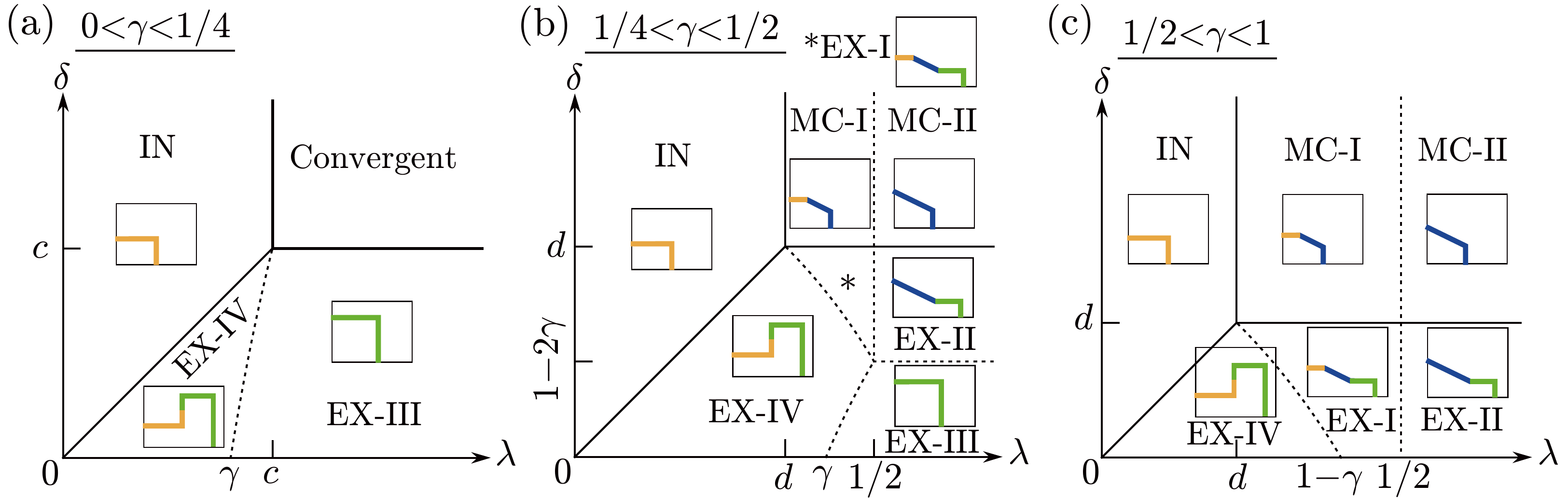}
 \caption{Phase diagrams including subphases of the divergent phase.
The structure of the phase diagram depends on the value of the growth rate 
 $\gamma$, i.e.  (a) $0<\gamma<\frac 1 4 $, (b) $\frac 1 4 <\gamma<\frac 1 2 $, (c) $ \frac 1 2 <\gamma < 1 $.
  The phase boundaries $c,d$ are given by $c = \frac{1-\sqrt{1-4\gamma}}{2}$ and $d = 1-\sqrt{\gamma}$.
  The boundaries between EX-I and EX-IV and between EX-III and EX-IV are given as 
  $ \lambda = 1- \frac{\gamma}{1-\delta} $ and $ \lambda =  \frac{\gamma}{1-\delta} $, respectively.
 }
\label{fig:sub}
\end{figure}

\begin{figure}
\begin{center}
\includegraphics[width=16cm]{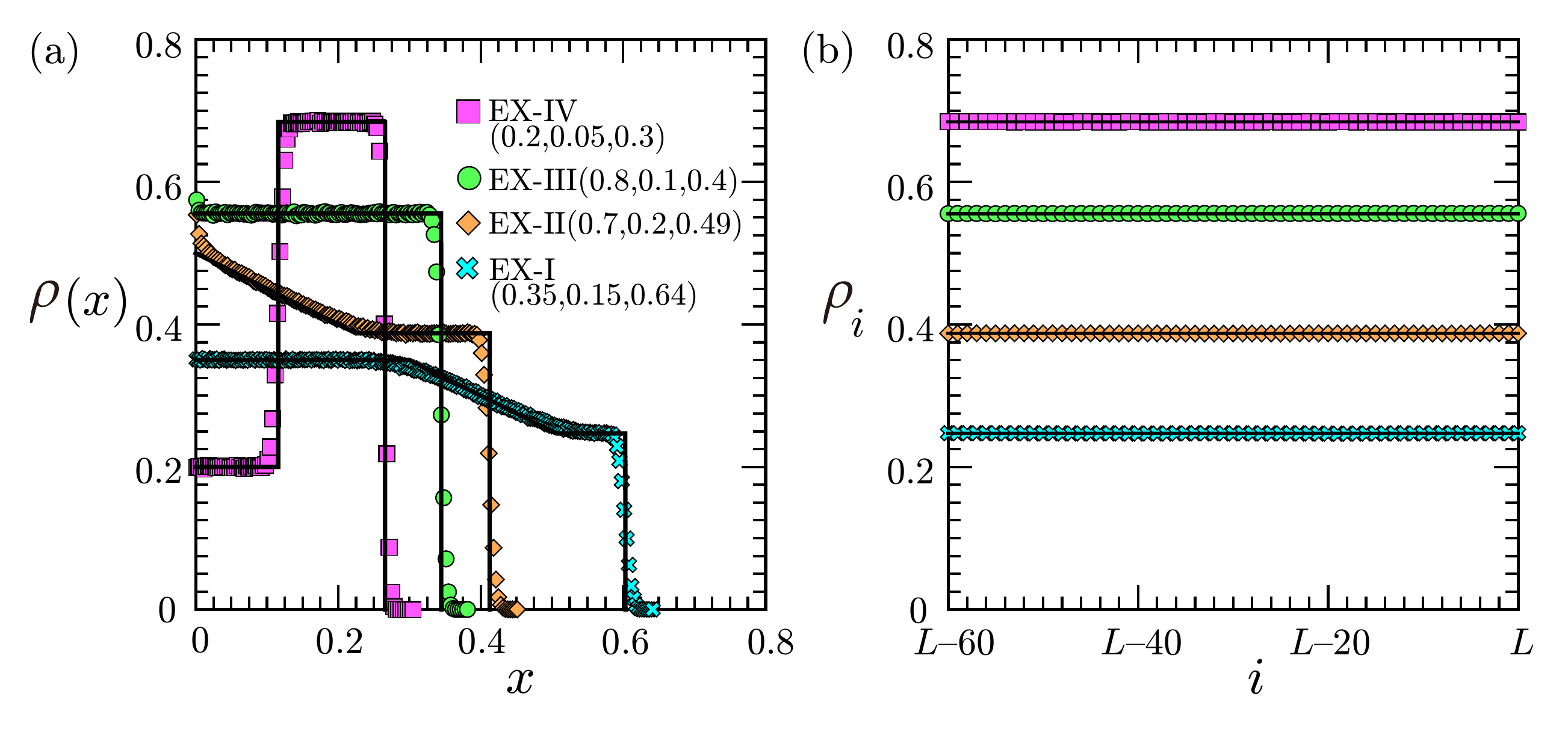}
\includegraphics[width=16cm]{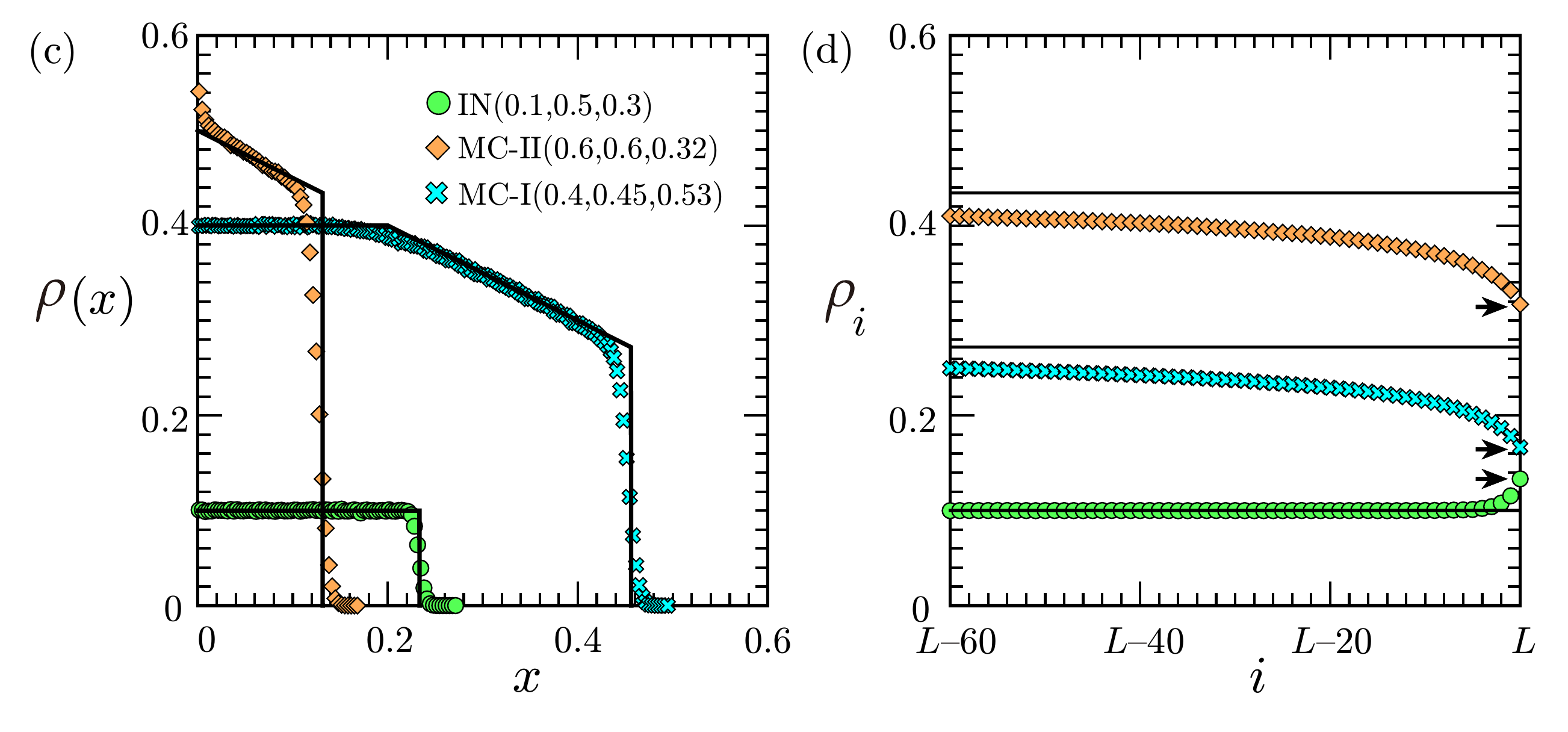}
\end{center}
\caption{
The macroscopic density profiles $\rho(x)$ in (a) EX, (c) MC and IN phases. We compare the results of Monte Carlo simulations (dots) with  the predictions by the hydrodynamic approach (lines). The microscopic density profiles $ \rho_i$ in the vicinity of the plus end $ (i=L,L-1,\dots) $  are also shown  for the (b) EX,  (d) MC and IN phases. We observe no boundary layer in the EX phase (b), but the true tip density $ \rho_L = \rho_+$ (indicated by arrows) is different from the macroscopic density $R$ (indicated by lines)  in the MC and IN phases (d) \cite{bib:MRF}, see Equations (\ref{eq:MC_rho+_v+}), (\ref{eq:IN_rho+_v+}) and (\ref{eq:R=lambda}). The chosen parameter values  $(\lambda,\gamma,\delta) $   are indicated in the panels (a) and (c),   and we used the same values and markers for (b) and (d), respectively.    As the initial condition we     chose $L=0$.  The results shown in (a) and (c) were obtained by   averaging $10\, 000$ snapshots at time $ t =15\, 000 $,   and for  (b) and (d) we averaged over the time window $10\, 000\le  t \le 15\, 000$   of  $10\, 000$ simulation samples.}
\label{fig:profiles}
\end{figure}

In order to determine the substructure of the EX phase, one has to check where each form (I,$\dots $, IV) of the density profile is realized. This can be done by evaluating $ \rho_\ell\gtrless \rho_r  $,  $v_s\gtrless 0$, $1-2\rho_+  \gtrless v_+$  and $1-2 \lambda \gtrless 0 $. The result is summarized in  Fig.~\ref{fig:sub}. Simulation results for the four sub-subphases are shown in Fig.~\ref{fig:profiles} (a,b).

\subsection{MC phase}

As we discussed in the previous subsection, a phase transition from the EX phase occurs, i.e.  the right plateau part $\rho_r = \rho_+$ (with Equation (\ref{eq:rho+EX})  of the rarefaction wave ($\lambda >  1-\sqrt \gamma$) vanishes, when $ \delta $  becomes larger than $1-\sqrt\gamma$, see Fig. \ref{fig:sub}.  Then the plus end {\it breaks into} the slope part of the rarefaction wave, and therefore the macroscopic density profile has the following structure: 
  
\begin{align}
\label{eq:DP-MC-I}
&   \rho(x)  |_ \text{MC-I}   =   
  \left\{\begin{array}{ll}
      \lambda  &       (0<x<1-2\lambda) ,\\
      \frac{1-x}{2} &       (1-2\lambda<x<v_+),   \\
  \end{array} \right.  \\
&   \rho(x)  |_ \text{MC-II} =    \frac{1-x}{2} \quad       (0<x<v_+) .   
\label{eq:DP-MC-II}
\end{align}
 The phase boundary between the two sub-subphases is given by $1-2\lambda=0$, i.e. $\lambda= \frac 1 2$.  Let us assume the tip density has a stationary value $\rho_+$. This true tip density is not the  limit of the macroscopic density profile (\ref{eq:R:=}), but related as
\begin{align}   R= \rho(v_+) =   \frac{1-v_+}{2}  . \end{align}
Here $v_+$ is the tip velocity which is written in terms of $\rho_+$ as   (\ref{eq:v_+=}).  Now we have obtained one relation between $\rho_+$ and   $R$: 
\begin{align}\label{eq:rhovrho+:1}  R =  \frac{1- ( \gamma   - \delta \rho_+)  } {2}  .\end{align}
In order to calculate $ \rho_+$ we need one more relation, which we can obtain from mass conservation, i.e. Equation  (\ref{eq:conservation}).   Solving them, we find  
\begin{align}
  \rho_+ =  \frac{(1-\sqrt\gamma)^2}{\delta}  ,\quad
 R = 1-\sqrt \gamma , \quad v_+ = 2\sqrt \gamma -1 .  
 \label{eq:MC_rho+_v+}
\end{align}
Simulation results for   the two sub-subphases  are shown in Fig.~\ref{fig:profiles} (c,d).

When $v_+ $ (\ref{eq:MC_rho+_v+})  becomes less than $ 1-2\lambda $ (i.e. $\lambda < 1-\sqrt\gamma $),  the slope part vanishes and we  enter the last subphase.

\subsection{IN phase}
As we discussed in the previous subsections, a phase transition from the EX phase occurs  when $\delta$ becomes larger than $\lambda$, and another one from the MC phase occurs when $\lambda$ becomes less than $ 1-\sqrt\gamma $, see Fig. \ref{fig:sub}. In the last subphase, the right plateau $\rho_r = \rho_+$ (\ref{eq:rho+EX}) of the shock profile (or the slope part of the rarefaction profile) vanishes, and only the left plateau remains, which was denoted as IN phase in \cite{bib:MRF}.  The macroscopic density profile is flat in the entire region as  
\begin{align}
    \rho (x) |_ \text{IN} = \lambda\quad  (0<x<v_+),
\end{align}
see  Fig.~\ref{fig:profiles}   (c).
 The tip density $\rho_+$ and the  velocity $v_+$ are given as 
\begin{align}
 \label{eq:IN_rho+_v+}  
   \rho_+ = \frac{\lambda(1-\lambda-\gamma)}{\delta(1-\lambda)} ,
  \quad v_+ = \gamma - \frac{\lambda(1-\lambda-\gamma)}{ 1-\lambda } 
\end{align}
from (\ref{eq:conservation}) with 
\begin{align}  \label{eq:R=lambda}  R=\lambda , \end{align}
see  Fig.~\ref{fig:profiles}   (d).

\section{Discussions}\label{sec:discuss}

In this work we studied a model for length regulation of  MTs by molecular motors  introduced in \cite{bib:MRF,bib:JEK}.  We restricted our discussion of the model to the case where the particle conservation is satisfied in the bulk,    i.e.   attachment and detachment of motors in the bulk are not considered.  This limit of the model has been suggested in \cite{bib:RMF}  and it has been argued that it is relevant for the length regulation by kinesin 5 due its high processivity. 

We gave an exact stationary solution of this model in the parameter regime where the average length of the filaments is finite. In this regime, we observe exponentially distributed filament lengths, similar to unregulated filaments.  The typical lengths of the MTs in this regime are rather short, except for parameter values that are at the limit to the divergent case. This result implies that length regulation of the filament is either generated by the bulk exchange of molecular motors, or by varying the influx of motors in  the vicinity of the transition from convergent to the divergent state. 

We also investigated the divergent phase by using a phenomenological approach and Monte Carlo simulations. The phenomenological approach enabled us to determine analytically a rich phase diagram. For infinitely growing MTs one can distinguish between  subphases where the entry (IN) and the exit (EX) induce macroscopic domains of constant densities. Additionally one observes a phase (MC) where the tip is connected to the slope part of a rarefaction wave. The distinction of these three phases had already been made in \cite{bib:MRF}. But in this work we explored the complete substructure of the divergent phase and possible shapes of density profiles by hydrodynamic approach,  and found good agreement with Monte Carlo simulations.

In view of our results it is of great interest to discuss the biological relevance of different regulatory mechanism of the system. In \cite{bib:GZH}  it was argued that processive motors are used for MT length regulation since they establish a direct link from the cell center to the periphery. For free filaments, however, our results indicate that length regulation is more robust, if the run length of molecular motors is finite. Therefore it would also be of great interest to see under which conditions the positioning of the microtubule organizing center would be possible in a finite volume, as it was observed for microtubules that are pulled by dynein motors attached to the boundaries in artificial cells \cite{bib:Laan}.

\section*{Acknowledgments}
 This work was supported by the SFB 1027.

\appendix
\section{Proof of stationarity}\label{sec:proof}

Here we prove that (\ref{eq:stationary-measure}) gives a stationary measure of the model.  The proof is very similar to the paper \cite{bib:DEHP}, where the matrix product ansatz was firstly introduced in the stochastic interacting particle system. To make this work self-contained, we write the proof in detail. 

The master equations concerning  $\tau= \emptyset, 0 $ and 1 are  
\begin{align}\label{eq:master:empty}
    \textstyle \frac d{dt}P_t(\emptyset) =& \delta P_t(1) - \gamma P_t(\emptyset), \\
 \label{eq:master:0}
    \textstyle \frac d{dt}P_t(0) =& \delta P_t(01) + \gamma P_t(\emptyset)   - (\lambda+\gamma) P_t(0) ,
\\
\label{eq:master:1}
    \textstyle \frac d{dt}P_t(1) =& \delta P_t(11) + \lambda P_t(0)  - ( \gamma + \delta ) P_t(1) .
\end{align}
For general configurations $\tau_1 \cdots \tau_L  $ with $L\ge 2$, the master equation is written as 
\begin{align} 
& \textstyle \frac d{dt} P_t  (    \tau_1 \cdots \tau_L )     \\
\label{eq:master:gen-lambda}
&= \lambda  (2 \tau_1 -1 )  P_t(  0\tau_2 \cdots \tau_L )   \\
\label{eq:master:gen-sum}
&   + \sum_{j=1}^{L-1}( \tau_{j+1} - \tau_j ) 
    P_t(\tau_1 \cdots \tau_{j-1} 1 0 \tau_{j+2} \cdots \tau_L )   \\ 
\label{eq:master:gen-gamma}
&  +  ( 1- \tau_L )  \gamma  P_t(\tau_1 \cdots \tau_{L-1} )
    - \gamma P_t(\tau_1 \cdots \tau_L )  \\ 
\label{eq:master:gen-delta}
&       + \delta P_t(\tau_1 \cdots \tau_L  1)
  -    \tau_L  \delta  P_t(\tau_1 \cdots \tau _{L-1}1 ). 
\end{align}

It is convenient to clearly distinguish the terms ``stationary measure'' and ``stationary distribution'' as defined in \cite{bib:Schi}.  A stationary measure is  a function $F$ in $S$ such that the right-hand sides of all the master equations become 0 by substitution $ P_t(\tau)  = F(\tau)  $. When $F$ is normalizable, i.e.  $ \sum_{ \tau\in S }  F(\tau)  $ is finite,  $ P (\tau) =   F(\tau) /  \sum_{ \tau\in S }  F(\tau)  $ gives a stationary distribution. On the other hand, when $ \sum_{ \tau\in S }F(\tau)  $ diverges, 
 the stationary measure loses its physical meaning \cite{bib:Schi}. 
 
Let us find a stationary measure in our case.  Without loss of generality we set $F(\emptyset ) =1 $. We soon notice that $  F(1) = \frac \gamma \delta  $  from (\ref{eq:master:empty}).  One wants to know the form of $  F( 0 ) $.   However we cannot find it by solving simultaneously  (\ref{eq:master:0}) and (\ref{eq:master:1}),  where apparently we need information of $   F(01) $ or  $  F(11)  $. Similarly, to know the form of $ F(\tau_1\dots \tau_L ) $  of   arbitrary configurations   $ \tau_1\dots \tau_L   $,  we need information of configurations with length $L+1$.   This fact prevents us from systematically finding a stationary measure, and thus an ansatz is required.

We now prove that the ansatz form (\ref{eq:stationary-measure}) actually gives a stationary measure.   Checking that  (\ref{eq:master:0}) $=0$ and  (\ref{eq:master:1}) $=0$ is not difficult.  For the general case,  the term (\ref{eq:master:gen-lambda}) becomes 
\begin{align}
  - (1-2\tau_1) \gamma^L  \langle W |  X_{\tau_2} \cdots X_{\tau_L }|V  \rangle   
\end{align}
thanks to (\ref{eq:algebra-left}).  Each summand of (\ref{eq:master:gen-sum}) becomes 
\begin{align}
&  +    (1-2\tau_j)
\gamma^L\langle W |  X_{\tau_1} \cdots X_{\tau_{j-1}} X_{\tau_{j+1}}  \cdots X_{\tau_L }|V  \rangle    \\
&  -   (1-2\tau_{j+1})
   \gamma^L\langle W |  X_{\tau_1} \cdots X_{\tau_j} X_{\tau_{j+2}} \cdots X_{\tau_L }|V  \rangle   
\end{align}
by using (\ref{eq:algebra-bulk}).  The summation of the last four terms i.e. (\ref{eq:master:gen-gamma})$+$(\ref{eq:master:gen-delta})  is calculated as 
\begin{align}
&     (1-2\tau_L)
\gamma^L\langle W |  X_{\tau_1} \cdots \cdots X_{\tau_{L-1} }|V  \rangle  , 
\end{align}
where (\ref{eq:algebra-right}) is used.  In total  $L$ cancels occur in   
(\ref{eq:master:gen-lambda})$+$(\ref{eq:master:gen-sum})$+$(\ref{eq:master:gen-gamma})$+$(\ref{eq:master:gen-delta}).   We have finished the proof.


\begin{thebibliography}{99}


\bibitem{bib:Alb} B~Alberts \textit{et al}, 2014 
\textit{Molecular Biology of the Cell}, 6th ed., Garland, New York    

\bibitem{bib:GZH} M~K~Gardner, M~Zanic, and J~Howard, 2013
\textit{Curr. Op. Cell Biol.}
\textbf{25}  14 

\bibitem{bib:EbbiS} M~Ebbinghaus and L~Santen, 2011
\textit{Biophys. J.}
\textbf{100}  832  

\bibitem{bib:GCRP}  M~L~Gupta, P~Carvalho, D~M~Roof and D~Pellman, 2006
\textit{Nat. Cell Biol.}
\textbf{8}  913   

\bibitem{bib:VHTHTH} V~Varga, J~Helenius, K~Tanaka, A~A~Hyman, T~U~Tanaka and J~Howard, 2006
\textit{Nat. Cell Biol.}
\textbf{8}  957 

\bibitem{bib:VLBDH} V~Varga, C~Leduc, V~Bormuth, S~Diez and J~Howard, 2009
\textit{Cell}
\textbf{138}, 1174  

\bibitem{bib:RMF} L~Reese, A~Melbinger  and E~Frey, 2011
\textit{Biophys. J.}
\textbf{101}  2190  

\bibitem{bib:JEK} D~Johann, C~Erlenk\"amper and K~Kruse, 2012
\textit{Phys. Rev. Lett.}
\textbf{108}  258103  

\bibitem{bib:MRF} A~Melbinger, L~Reese and E~Frey, 2012
\textit{Phys. Rev. Lett.}
\textbf{108}  258104   

\bibitem{bib:RMFa} L~Reese, A~Melbinger and E~Frey, 2014
\textit{Interface Focus.}
\textbf{4} 20140032    

\bibitem{bib:Schu}   G~M~Sch\"utz, 2001
\textit{Exactly Solvable Models for Many-Body Systems Far from Equilibrium}
 in \textit{ Phase Transitions and Critical Phenomena vol 19.},
  C~Domb and J~L~Lebowitz Ed.,   Academic Press, San Diego    
  
\bibitem{bib:MG}  C~T~MacDonald J~H~Gibbs  and  A~C~Pipkin, 1968
\textit{Bioplymers}
\textbf{6}  1  

\bibitem{bib:DEHP}  B~Derrida, M~R~Evans, V~Hakim and V~Pasquier, 1993
\textit{J. Phys. A: Math. Gen}
\textbf{26}  1493  

\bibitem{bib:SEPR} K~Sugden, M~R~Evans, W~C~K~Poon  and N~D~Read, 2007
\textit{Phys. Rev. E}
\textbf{75}  031909  

\bibitem{bib:SE} K~Sugden and M~R~Evans, 2007
\textit{J. Stat. Mech.}, P11013 

\bibitem{bib:EvanS} M~R~Evans and K~E~P~Sugden,   2007
\textit{Physica A}
\textbf{384}  53  

\bibitem{bib:DMP} S~Dorosz, S~Mukherjee  and T~Platini, 2010
\textit{Phys. Rev. E}
\textbf{81}  042101  

\bibitem{bib:Ari} C~Arita, 2009
\textit{Phys. Rev. E}
\textbf{80}  051119   

\bibitem{bib:AS} C~Arita and A~Schadschneider, 2012
\textit{J. Stat. Mech.}
  P12004  

\bibitem{bib:dGF} J~de~Gier and  C~Finn, 2014
\textit{J. Stat. Mech.}
  P07014 

\bibitem{bib:RMFb} L~Reese,  A~Melbinger and E~Frey, 2015
arXiv:1505.01219    

\bibitem{bib:L} A~L\"uck, 2014 Master thesis, Saarland University  

\bibitem{bib:H} O~J~Heilmann, 2004
\textit{J. Stat. Phys.}
\textbf{116}  855   
 
\bibitem{bib:KRB} P~L~Krapivsky, S~Redner and E~Ben-Naim, 2010
\textit{A Kinetic View of Statistical Physics}, Cambridge University Press, Cambridge  

\bibitem{bib:BJHK}  R~A~Blythe, W~Janke, D~A~Johnston and R~Kenna, 2004
\textit{J. Stat. Mech.}  P10007 

\bibitem{bib:BE} R~A~Blythe and M~R~Evans,  2007
\textit{J. Phys. A: Math. Gen}
\textbf{40}  R333 

\bibitem{bib:Schi} R~B~Schinazi, 1999
\textit{Classical and Spatial Stochastic Process}, Birkh\"auser, Boston  

\bibitem{bib:Laan} L~Laan \textit{et al.},  2012
\textit{Cell}
\textbf{148} 502  


\end{thebibliography}
\end{document}